# Time varying gratings model Hawking radiation

Simon A. R. Horsley[1] and John B. Pendry[2]

[1] School of Physics and Astronomy, University of Exeter, Stocker Road, Exeter EX4 4QL, UK

[2] The Blackett Laboratory, Department of Physics, Imperial College London, London SW7 2AZ, UK

**Email:** j.pendry@imperial.ac.uk

## Abstract

Diffraction gratings synthetically moving at trans-luminal velocities contain points where wave and grating velocities are equal. We show these points can be understood as a series of optical event horizons where wave energy can be trapped and amplified, leading to radiation from the quantum vacuum state. We calculate the spectrum of this emitted radiation, finding a quasi-thermal spectrum with features that depend on the grating profile, and an effective temperature that scales exponentially with the length of the grating, emitting a measurable flux even for very small grating contrast.

## Significance Statement

Electromagnetic radiation is strongly influenced by structure in the permittivity and permeability. Structures modulated in space and time can appear to move with speeds at or above the speed of light and produce remarkable effects of increasing interest theoretically spurred by the possibility of experimental realization. We take theory into the quantum domain identifying singularities in transluminal structures analogous to the Schwarzschild singularity in the gravitational metric of a black hole. We show that these singularities will spontaneously radiate photon pairs, analogous to Hawking radiation. A temperature can be associated with the radiation which depends exponentially on the length of time for which the structure in is motion. In addition to spontaneous radiation stimulated emission is produced by incident radiation.

**Introduction**

Space-time modulations of the refractive index can lead to unusual wave propagation, where common assumptions such as frequency conservation no longer hold. This has been known since the early work of Morgenthaler [1] and Mendonça [2] but until recently there was little possibility of any experiments in this area. Yet space-time varying material parameters can now be implemented using non-linear optical effects [3] and acoustic signal processing [4], sparking an ongoing deepening theoretical understanding [5]. These new developments have included the discovery of Fresnel drag in synthetically moving media [6], the theory of homogenization for space-time varying materials [7], energy re-direction through varying anisotropy ("temporal aiming") [8], polarization selective wave amplifiers [9], and temporal anti-reflection coatings [10].

Diffraction gratings are an interesting special case [11,12,13,14], and a new gain mechanism has been discovered in these temporal gratings [15,16,17]. A moving grating profile, $n(x - c_g t)$, traps wave energy at points where the grating velocity $c_g$ equals the local wave speed $c = c_0/n$. As the wave moves through the grating, its energy is concentrated at an exponential rate around these accumulation points such that a continuous wave input is transformed – via a linear equation - into a series of extremely short, intense electromagnetic pulses (this process is shown in the final panel of Fig. 1).

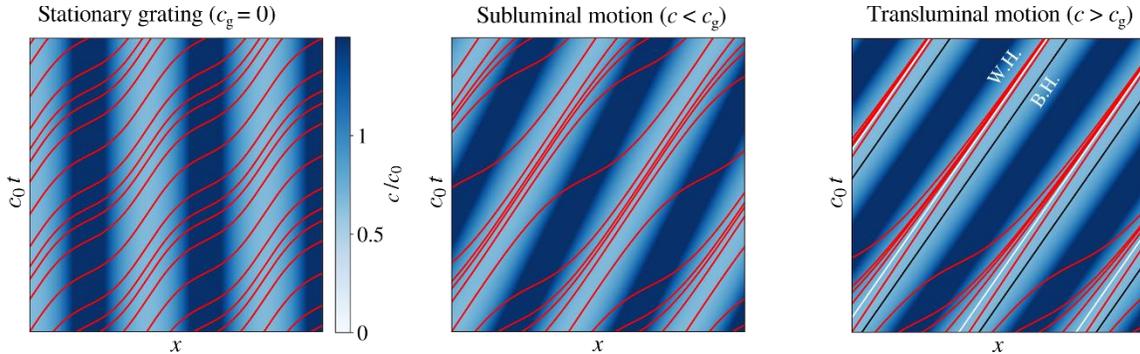

**Figure 1**: *Event horizons in a moving grating*: (left) a collection of rays (red) propagates through a stationary diffraction grating with variable wave speed $c$, indicated via the blue shading. (centre) A grating moving with speed $c_g$, tends to locally concentrate or disperse rays around points where $|c_g - c|$ is minimal. (right) A transluminal grating contains points where $c_g = c$ through which rays cannot propagate. At these points the rays are either highly concentrated (white lines) or dispersed (black lines), equivalent to white (W.H.) and black hole (B.H.) event horizons, respectively.

As found in [17], there is a similarity between wave propagation close to one of these accumulation points, and close to a space-time event horizon. Investigating this equivalence, and the implied Hawking-like radiation is the subject of this paper. At points where the grating and wave velocities are equal, energy either accumulates $(dc/dX < 0)$, or disperses $(dc/dX > 0)$, as shown on the right of Fig. 1. The wave is drawn away from a point where $dc/dX < 0$, without ever crossing this point, analogous to outwards propagation from a *black hole* event horizon. An accumulation point is the time reverse of this, and is thus equivalent to inwards propagation towards a *white hole* event horizon: a point that the wave cannot enter. Transluminal space-time gratings can therefore be understood as an effective space-time geometry for light, containing an alternating sequence of optical black and white hole event horizons. This is an extreme instance of transformation optics - the established equivalence between electromagnetic materials and space-time geometry [18,19] - and connects to earlier work investigating optical event horizons in non-uniformly moving media [20,21].

Galilean transformation, $X = x - c_g t$, to a co-moving frame in which the grating is stationary reveals a singularity.

$$\varepsilon_{mov} = \frac{\varepsilon(X)}{1 - c_g^2/c^2(X)}, \quad \mu_{mov} = \frac{\mu(X)}{1 - c_g^2/c^2(X)}, \quad \xi_{mov} = -\frac{\varepsilon(X)\mu(X)c_g}{1 - c_g^2/c^2(X)}, \quad (1)$$

where $1/c^2(X) = \varepsilon(X)\mu(X)$ and $\xi_{mov}$ couples electric and magnetic fields. We refer the reader to these earlier papers for derivation of the constitutive relations in the new frame [6,7]. The singularity formed when $c(X) = c_g$ is of the same order as seen when light encounters the event horizon in an black hole and is the basis for the analogy we make.

Given this series of optical event horizons in a moving grating, each can be expected to emit Hawking radiation from the vacuum state [22]. Although analogue Hawking radiation has been predicted and observed for isolated optical and sonic event horizons [23,24,25,26], here we calculate its spectrum for a time varying grating confined to a fixed length window within an otherwise homogeneous medium. This approach to observing analogue Hawking radiation from multiple optical horizons has not been considered before, and we find the effective temperature of the emitted radiation scales exponentially with the length of the grating. Thus, even low contrast transluminal gratings can emit a large amount of radiation from the vacuum state.

**Maxwell's equations tell a quantum story**

Like Schrödinger's equation Maxwell's equations describe the physics of a particle as a wave. The absence of Planck's constant due to the zero mass of the photon hides the essentially quantum nature of these equations which already foretell many quantum effects. We begin with a transluminal grating of spatial period $2\pi/g$.

$$\varepsilon(gx - \Omega t) = \mu(gx - \Omega t) = \varepsilon_b + 2\alpha \cos(gx - \Omega t) \quad (2)$$

Eq. (2) defines a PT symmetric system and the absence of back scattering implied by the constant impedance of the model yields dispersion relationships with real frequencies and real wave vectors. This model has been extensively investigated and much is known about its classical behaviour [15,16,17]. It is especially interesting when the local velocity of light at some point within the grating equals the grating velocity, $c_g = \Omega/g$, which then acts as an accumulation point for energy. An intense peak of radiation forms at this point whose energy grows exponentially with length of grating. This despite the absence of band gaps. The existence of such points defines a transluminal grating.

It is known from previous work [15] that the pulses have the form,

$$f'(x)e^{-i\omega c_0^{-1}f(x)} \quad (3)$$

where $\omega$ is the frequency of the incident radiation. To a high degree of accuracy $f'(x)$ has a Lorentzian form,

$$f(x) = i\ln\frac{xg + 2i\gamma}{xg - 2i\gamma}, \quad f'(x) = \frac{4\gamma}{x^2 g^2 + 4\gamma^2} \quad (4)$$

Analytic theory [1] gives the value of $\gamma$ as,

$$\gamma = e^{-2\alpha g d} \quad (5)$$

Through performing a Fourier transform on Eq. (3) and concentrating on negative frequencies we calculate the spectrum of emerging radiation,

$$F(\tilde{k},n,n') = \frac{1}{2\pi}\int_{-\infty}^{+\infty} f'(x) e^{i(\tilde{k}+ng)f(x) - i(\tilde{k}+n'g)x} dx$$

$$= +2 e^{-\pi i(\tilde{k}/g+n)} \frac{\sin(\pi(\tilde{k}/g+n))}{\pi(\tilde{k}/g+n)} (\tilde{k}+n'g) \times \int_{+2\gamma/g}^{+\infty} e^{+(\tilde{k}+n'g)z} \left(\frac{zg-2\gamma}{zg+2\gamma}\right)^{(\tilde{k}/g+n)} dz \qquad (6)$$

where $c_0(\tilde{k}+ng)$ is the incident frequency which we assume to be positive and $c_0(\tilde{k}+n'g)$ is one of the exit frequencies which we assume to be negative. We define $\tilde{k}$ as the input wave vector modulo $g$. For a long grating ($gd \gg 1$) the spatial/temporal transmission function (3) is highly localised ($\gamma \ll 1$), we can approximate the integral in (6) to give,

$$\lim_{n' \to \infty} |F(\tilde{k},n,n')|^2 = 4 \frac{\sin^2(\pi(\tilde{k}/g+n))}{\pi^2(\tilde{k}/g+n)^2} e^{+(\tilde{k}/g+n')4\gamma} \qquad (7)$$

where we have assumed $n' \gg \gamma^{-1}$. Which shows that the energy content of the mode emerging from the grating is proportional to,

$$e^{+4\gamma(\tilde{k}/g+n')} = e^{-\hbar|\omega_{n'}|/(k_B T_H)} \qquad (8)$$

where we have remembered that $(\tilde{k}/g+n') < 0$ and have interpreted the exponential dependence on photon energy in terms of an effective Hawking temperature, $T_H$. From our definition of $\gamma$ (5) we see that this temperature increases exponentially with the length of grating,

$$T_H = \frac{\hbar g c_0}{4 k_B} e^{2\alpha g d} = \frac{\hbar \Omega}{4 k_B} e^{2\alpha g d} \qquad (9)$$

These equations derived from the classical limit already hint that something special happens at negative frequencies. Next we further probe the significance of the negative.

Recently conservation laws in this system have been studied [15] and it was found that energy is not always conserved but instead rigorous conservation applies to a pseudo photon number, $\tilde{N}$, defined as,

$$\tilde{N} = \sum_n \frac{U_n}{\hbar \omega_n}, \quad N = \sum_n \frac{U_n}{|\hbar \omega_n|} \qquad (10)$$

where $U_n$ is the flow of energy in the $n$th mode and $\omega_n$ the frequency of that mode. $\tilde{N}$ sits in contrast to the true photon number, $N$, which recognises that frequencies may be negative but photon energies are always positive. The two expressions do not conflict if transitions are confined to frequencies of the same sign, but if photons make a transition that flips the sign of the frequency there is a conflict, and it follows that,

$$N - \tilde{N} = 2 \qquad (11)$$

Not only have we removed a negative energy photon from $\tilde{N}$ but we have added a second one with positive energy to $N$. Thus Maxwell's equations already inform us that our system is capable of creating qubits. The presence of negative frequencies in the transmitted spectrum implies that qubits have been formed in an essentially quantum process

Constructing a Hamiltonian for the system [15] confirms this conclusion,

$$\hat{H} = \frac{\hbar}{2} \sum_{bb'} \sum_{nn'>0} |\omega_{bn}\omega_{b'n'}|^{1/2} \left(\varepsilon'^{-1}\right)_{bnb'n'} \left[\hat{a}_{bn}\hat{a}^\dagger_{b'n'} + \hat{a}^\dagger_{bn}\hat{a}_{b'n'}\right]$$
$$+ \frac{\hbar}{2} \sum_{bb'} \sum_{nn'<0} |\omega_{bn}\omega_{b'n'}|^{1/2} \left(\varepsilon'^{-1}\right)_{bnb'n'} \left[\hat{a}_{bn}\hat{a}_{b'n'} + \hat{a}^\dagger_{bn}\hat{a}^\dagger_{b'n'}\right]$$
(12)

Photons injected with positive frequencies cannot pass the ± threshold in frequency but instead populate frequencies of opposite sign by stimulated emission of qubits the two members of which straddle the divide in sign and thereafter lead independent lives. Spontaneous emission can be thought of as emission stimulated by vacuum fluctuations and this is the approach we take here.

Consider a finite section of grating. Radiation of unit amplitude incident from the left with frequency is transmitted by the grating into a range of frequencies with amplitudes so that the energy flux at each transmitted frequency is,

$$U(\omega + n'\Omega) = U_{inc} |T(\omega + n'\Omega, \omega)|^2$$
(13)

where $U_{inc}$ is the incident flux. Now our strategy is clear: we substitute $U_{inc} = c_0 \hbar \omega / 2$, corresponding to the right-going energy flux in the vacuum state, we shall be able to calculate the spontaneous emission from the grating.

There is however one fly in the ointment. Classical calculations are equivalent to real photons: the first term in Eq. (12) is the quantum counterpart of classical frequency conversion, allowing photons to climb up and down the frequency scale. This process is not available to the ground state because the annihilation operators in the first term of (12) give a null result. Only the second term in (12) survives, which represents the photon creation process. Therefore, if our classical calculation injects a positive frequency wave to mimic the vacuum state, positive frequencies emerging from the grating will be polluted by these frequency shifts which will spread the spectrum over all positive frequencies.

However as just observed, the first term in (12) cannot couple a positive frequency input to negative frequencies. Negative frequencies can be populated only by stimulated emission and therefore the negative frequency flux will give a true representation of the processes available to vacuum fluctuations. Furthermore if we integrate over all positive frequency fluctuations the negative frequency fluxes will be symmetrical with the other half of the qubits created at positive frequencies. Adding in the contribution of negative frequency incident fluctuations gives a factor of two. The final formula for spontaneous quantum radiation is then,

$$U(\omega + n'\Omega) = 2c_0 \sum_n \frac{\hbar(\omega + n\Omega)}{2} |T(\omega + n'\Omega, \omega + n\Omega)|^2,$$
$$\omega + n\Omega > 0, \quad \omega + n'\Omega < 0$$
(14)

Next we ask how many photon pairs emerge from a length of grating, in the limit of a long grating. We can calculate this number from Eq. (6) and show that,

$$\lim_{t \to \infty} N_{phot} = \lim_{t \to \infty} \sum_{n'=-1}^{-\infty} \frac{\left|\tilde{F}(\tilde{k},n,n')\right|^2}{\hbar c_0 \left|\tilde{k}+n'g\right|}$$

$$= -\frac{4}{\hbar c_0} \frac{\sin^2\left(\pi\left(\tilde{k}/g+n\right)\right)}{g\left(\tilde{k}/g+n\right)^2} \int_{+2}^{+\infty} \frac{1}{(z'+z'')^2} \left(\frac{z'-2}{z'+2}\frac{z''-2}{z''+2}\right)^{(\tilde{k}/g+n)} dz'dz'' \quad (15)$$

where we have assumed a fixed incident wave of frequency $\omega = c_0(\tilde{k}+ng)$. The expression is clearly independent of $\gamma$ and therefore of the length of grating. We computed Eq. (15) numerically and the result we plot in Fig. 2.

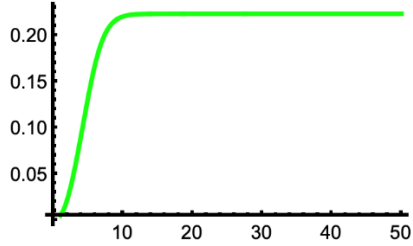

**Figure 2**. Plot of the integral in Eq. (15) versus $-\log \gamma(d)$ for $\tilde{k}=0.5$, $n=1.0$, $g=1.0$, $\alpha=0.05$,

which shows the photon number saturating with length of the grating.

Saturation of the curve indicates that the number of pairs created in a length of grating is always finite. However at the same time a process of amplification is taking place as the photons created climb a frequency ladder. This can be seen by summing over all the field intensities in the emerging modes,

$$\lim_{d \to \infty} \sum_{n'=-1}^{-\infty} \left|\tilde{F}(\tilde{k},n,n')\right|^2 = \frac{\sin^2\left(\pi\left(\tilde{k}/g+n\right)\right)}{\left(\tilde{k}/g+n\right)^2} \frac{e^{+2\alpha gd}}{g} \quad (16)$$

The ladder-climbing process rapidly depletes frequencies in the vicinity of zero, the hot spot for pair creation, as photons flee to large positive and negative frequencies. Creation of qubits is largely confined to low values of $n$.

**Fully quantum mechanical calculation**

The semi-classical argument presented in the previous section can be justified using a fully quantum mechanical calculation. We use the Heisenberg picture [27], where the time variation of the system is imprinted in the field operators, rather than the state vector. As discussed above, classically we can characterize a windowed grating in terms of a two-time transmission function $T(t,t')$, e.g. for left to right propagation,

$$E_f(t) = \int_{-\infty}^{+\infty} T(t,t') E_i(t') dt' \quad (17)$$

Eq. (17) captures the linear relation between the time dependent outgoing field $E_f(t)$ at $x = d$ in terms of the incoming field $E_i(t')$, at $x = 0$, which encodes the solution to Maxwell's equations within the grating. Given that in the Heisenberg field operators obey the classical Maxwell equations, e.g. $\partial \tilde{E}/\partial x = \mu_0 \, \partial \tilde{H}/\partial t$, and the same boundary conditions, Eq. (17) also holds for the field operators,

$$\hat{E}_f(t) = \int_{-\infty}^{+\infty} T(t,t') \hat{E}_i(t') dt' \tag{18}$$

Eq. (18) now connects the vacuum state on the input side of the grating – defined the same way as in free space, as the zero eigenstate of the incoming wave annihilation operators $\hat{a}_i |0\rangle = 0$ - to the corresponding out-going state, which in general will not be the zero photon state. To find the relationship between the creation and annihilation operators on the two sides of the grating, we decompose the field operators as integrals over frequency,

$$\hat{E}_{f,i}(t) = i \int_0^{+\infty} d\omega \sqrt{\frac{\hbar \omega}{2\varepsilon_0 c_0}} \left[ \hat{a}_{f,i}(\omega) e^{-i\omega t} - \hat{a}^\dagger_{f,i}(\omega) e^{+i\omega t} \right] \tag{19}$$

Substituting this expansion of the input and output field operators into Eq. (18), we can relate the creation and annihilation operators on the two sides of the grating,

$$\hat{a}_f(\omega') = \int_0^{+\infty} d\omega \sqrt{\frac{\omega}{\omega'}} \left[ \hat{a}_i(\omega) T(\omega',\omega) - \hat{a}^\dagger_i(\omega) T(\omega',-\omega) \right] \tag{20}$$

Where $T(\omega',\omega)$ is the double Fourier transform of the two-time transmission function encountered in Eq. (14) of the previous section. Eq. (20) determines the vacuum radiation emitted from the grating, and is an integral form of a Bogoliubov transformation [28]. Importantly, when the grating mixes positive and negative frequencies, $T(\omega',-\omega) \neq 0$, the *annihilation* operators $\hat{a}_f$ for out-going modes depend on the *creation* operators $\hat{a}^\dagger_i$ for incoming modes. This means that the number of outgoing photons in the vacuum state, counted using $\langle 0|\hat{a}^\dagger_f \hat{a}_f |0\rangle$, is non-zero because the vacuum state is not reduced to zero by $\hat{a}_f$. Note that the presence of optical event horizons in the grating is crucial to this emission process. The local frequency of a wave propagating through the grating [15], $\omega/(1 - c_g/c)$, changes sign across each point where $c_g = c$.

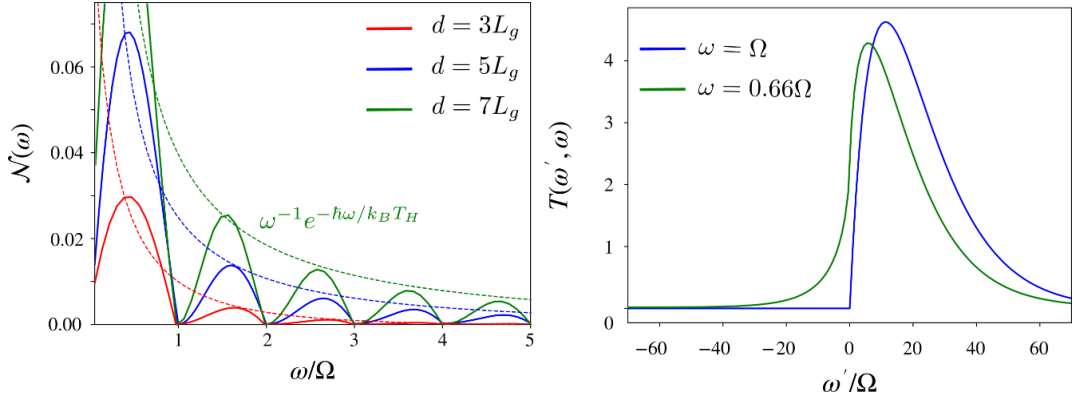

**Figure 3**: *Numerical calculation of vacuum emission*: (left) Emitted photon number density per grating period calculated from Eq. (24) for $d = 3, 5, 7 \times L_g$. For this calculation we used $\alpha_\varepsilon = \alpha_\mu = 0.05$, $c_0 = 1.0$, $\Omega = 1.0$, $g = \Omega$. Emission is peaked in the range $0 < \omega < \Omega$ and vanishes at multiples of the grating frequency. The dashed lines are the curves $\omega^{-1} e^{-\hbar \omega / k_B T}$ scaled to pass through the maximum of the spectrum, and with the effective temperature given by Eq. (9). (right) Numerical calculation of two frequency transmission function $T(\omega, \omega')$ for $d = 5 L_g$ and input frequencies $\omega = \Omega$ (blue) and $\omega = 0.66 \Omega$ (green). As anticipated, for an input frequency equal to an integer multiple of the grating frequency the transmission function does not couple positive and negative frequencies.

To quantify the vacuum radiation, we calculate the average flow of electromagnetic energy emerging from the grating. In our simplified case the materials are impedance matched and we can write the Poynting vector operator in terms of the square of the electric field $\hat{S}_x = \hat{E}\hat{H} = \eta_0^{-1} \hat{E}^2$. Applying the operator expansion (19) and the transformation of the creation and annihilation operators (20), the total emitted energy $U$ is,

$$U = \int_{-\infty}^{+\infty} \langle 0|\hat{S}_x(t)|0\rangle dt = \int_0^{+\infty} d\omega \hbar \omega \left[ \langle 0|\hat{a}_f^\dagger \hat{a}_f(\omega)|0\rangle + \frac{1}{2}\delta(0) \right]$$

$$= \int_0^{+\infty} d\omega \hbar \omega \left[ \int_0^{+\infty} d\omega' \frac{\omega'}{\omega} |T(\omega', -\omega)|^2 + \frac{1}{2}\delta(0) \right] \tag{21}$$

The infinite term $\delta(0)/2$ on the right of Eq. (21) is the zero photon contribution to the energy flow, which is cancelled as soon as we include the right-to-left travelling waves in the calculation: from here on we drop this divergent term.

Eq. (21) holds for any *linear* medium. For the grating illustrated in Fig. 1, the transmission function is periodic in time, which simplifies the general expression. Writing the transmission function, $T$, for an $N$ period grating as a sum over the transmission, $\tilde{T}$ through one period, displaced by multiples of the grating temporal period $\Delta_g$.

$$T(\omega,\omega') = \sum_{n=0}^{N-1} e^{i(\omega'-\omega)n\Delta_g} \tilde{T}(\omega,\omega')$$
$$= e^{i(N-1/2)(\omega'-\omega)\Delta_g/2} \frac{\sin[(\omega'-\omega)N\Delta_g/2]}{\sin[(\omega'-\omega)\Delta_g/2]} \tilde{T}(\omega,\omega') \quad (22)$$

For large $N$ the ratio of sines on the right of Eq. (22) oscillates rapidly with a significant contribution only for frequencies $\omega' = \omega + N\Omega$ where $\Omega = 2\pi/\Delta_g$. Substituting the transmission function (22) into our expression for the emitted energy (21) and applying the identity

$$\lim_{N\to\infty} \frac{1}{\pi N} \frac{\sin^2(Nx)}{\sin^2(x)} = \sum_{n=-\infty}^{+\infty} \delta(x+n\pi) \quad (23)$$

the emitted energy per grating temporal period is given by,

$$\frac{U}{N} = \frac{1}{\Delta_g} \int_0^{+\infty} d\omega \hbar\omega \sum_{\omega>0} \frac{\omega_n}{\omega} |\tilde{T}(\omega,-\omega_n)|^2 = \int_0^{+\infty} d\omega \hbar\omega \mathcal{N}(\omega) \quad (24)$$

Where $\omega_n = \omega + n\Omega$ and $\mathcal{N}$ is the number density of emitted photons per grating period. Eq. (24) shows that the emitted energy at each frequency $\omega$ is proportional to the square of the transmission function multiplied by the ratio of the photon energies and summed over all negative input frequencies $\omega_n$, as found in Eq. (14) of the above semi-classical argument.

As explained in the previous section, the problem of calculating the emitted spectrum of photons is reduced to finding the portion of the two-frequency transmission function that mixes positive and negative frequencies. In Fig. 3 we evaluate the spectrum given in Eq. (24), using a numerical calculated transmission function. The figure shows that the photon density $\mathcal{N}(\omega)$ oscillates as it decays exponentially with frequency. As we anticipated in our semi-classical analysis, the left panel of Fig. 3 shows that the emission spectrum is null at frequencies that are an integer multiple of the grating frequency. This is a feature of the cosine grating, arising because it yields a tridiagonal operator, coupling only to frequencies $\omega \pm \Omega$. Additionally, the dashed lines in this panel show that the exponential decay of the spectrum with frequency is well approximated by the effective temperature identified in Eq. (9). Our analysis is further confirmed in the right hand panel of the figure, which shows that when the input frequency is an integer multiple of the grating frequency, the transmission function no-longer mixes positive and negative frequencies. Sloan et al have proposed a related method for the generation of entangled pairs [29].

**Stimulated vacuum radiation**

Transluminal gratings with optical wavelength periods spontaneously emit photons with an eV or so of energy and for the numbers we show to be created these are easily detectable. The difficulty is that producing such short period gratings is a big experimental challenge even though we are demanding only a few percent modulation. Moving to long spacings, the experimental challenge of manufacture becomes more feasible, but the energy output scales inversely as the spacing and soon becomes undetectable. Perhaps the most easily implemented experiment would be to switch to acoustic waves, but there too detectability of spontaneous emission suffers from an insurmountable intensity problem.

However, the same emission process of qubit creation is also evident in stimulated emission, where the grating is probed using a classical input. As we have argued above, a positive frequency input can only stimulate emission at negative frequencies by the creation of qubits, even though they may not be individually detectable. We shall show that a substantial fraction of the input energy can be converted to negative frequencies.

Although negative frequencies do not show up as such in a spectrometer, they are easily identified if the stimulating frequency is not an integer multiple of $\Omega/2$. Suppose that the positive input frequency is $\tilde{\omega}$ modulo $\Omega$. Then all positive frequencies would show as $\tilde{\omega}$ modulo $\Omega$ and all negative frequencies as $|\Omega - \tilde{\omega}|$ modulo $\Omega$, distinguishing them from the positive frequencies.

Taking Eq. (14) as the starting point, in Fig. 4 we plot $\left|F\left(\tilde{k}=0.75,1,n'\right)\right|^2$ against $n'$.

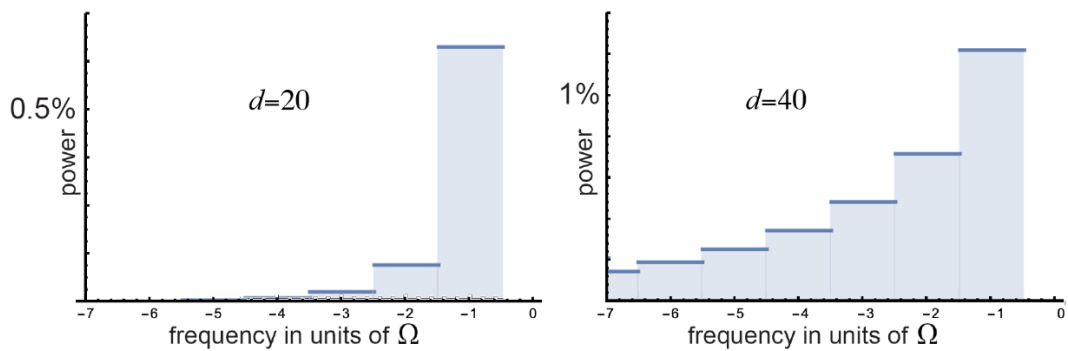

**Figure 4**. *Calculation of the fraction of input power delivered at negative frequencies*: Plot of $\left|F\left(\tilde{k}=0.75,1,n'\right)\right|^2$ against $n'$ showing the fraction of incident radiation converted to stimulated emission at each negative frequency. The parameters are for left $d=20$, for right $d=40$, and for both $\alpha=0.05$, $g=1$, $\tilde{k}=0.75$, $n=1$.

The total flux of power summed over all negative frequencies is 0.7% (left) and 4.4% (right) showing that even for relatively long-period gratings, the power available for detection is substantial and unlikely to be lost in background noise. We might envisage a transmission line experiment operating at GHz frequencies where external modulation of the components is used to create the required luminal grating.

**Summary and Conclusions**

A transluminal diffraction grating traps and amplifies radiation at points where the local wave and grating velocities are equal. For a simple cosine modulation of the refractive index these points occur in pairs, one of which is such an accumulation point for wave energy, whereas the other is the time reverse of this, from which electromagnetic radiation is extinguished. We have demonstrated the equivalence of these points to white and black hole event horizons respectively and calculated the associated Hawking radiation: the photon emission that occurs when the vacuum state is incident onto the grating.

In contrast to previous work, which typically finds a thermal spectrum through applying a WKB-like approximation to an isolated event horizon, we have calculated the two-frequency transmission function $T(\omega,\omega')$ for the windowed multi-horizon geometry shown in Fig. 1. This set up could be experimentally realised via illumination of a non-linear medium with an interference pattern between chirped beams. The general formulae for the emission spectra calculated in Eqns. (14) and (24) shows that the vacuum radiation is determined by the classical amplitude for radiation to be shifted from positive to negative frequencies (or the reverse) after transmission through the grating.

Despite neglecting material dispersion and assuming impedance matching, we found that the emitted radiation from transluminal gratings has a spectrum that is quasi-thermal, with features that depend on the grating profile. For instance Fig. 3 shows that while the photon number density decays exponentially with frequency, it is zero at multiples of the grating frequency, $\Omega$. This can be understood in terms of the Fourier representation of a cosine, which is tridiagonal and couples the wave only to frequencies that differ by $\pm\Omega$. Emission at multiples of the grating frequency thus requires the wave to pass through zero frequency where the coupling to the grating vanishes.

Interestingly the effective temperature of the emitted radiation is given by Eq. (9), and scales exponentially with the grating length. This indicates a large flux of electromagnetic vacuum radiation for a low contrast grating with a window containing only tens of grating periods. In reality this exponential dependence on the grating length is curtailed by material dispersion, as the emitted radiation spreads over an ever larger range of frequencies. However, our results are not restricted to electromagnetism and analogous phenomena should be present in pressure acoustics or elasticity, where material dispersion may be reduced.

Although the quantum vacuum radiation can be identified through coincidence photon counting of oppositely circularly polarized photons, it may be experimentally challenging to reduce the temperature such that the pure vacuum state is incident on the grating. Nevertheless, classical stimulated emission experiments can reveal the same amplification process and positive to negative frequency conversion. This makes acoustics a promising platform for investigating these effects, connecting to recent work demonstrating amplification of acoustic waves through positive to negative frequency conversion [30,31].

## Acknowledgments

SARH acknowledges the Royal Society and TATA for financial support through grant URF\R\211033.
 JBP. acknowledges funding from the Gordon and Betty Moore Foundation.